\documentclass{article}
\usepackage{dcase2023,amsmath,graphicx,url,times,booktabs, tabularx}
\usepackage{multirow, amssymb}

\title{Pretraining Representations for Bioacoustic Few-shot Detection using Supervised Contrastive Learning}

\name{Ilyass Moummad$^{1}$\thanks{This work was co-funded by the AI@IMT program and the company OSO-AI.},
      Romain Serizel$^{2}$, 
      Nicolas Farrugia$^{1}$, 
      }
\address{$^1$ IMT Atlantique, Lab-STICC, UMR CNRS 6285, Brest, France,\\ 
        \{ilyass.moummad, nicolas.farrugia\}@imt-atlantique.fr\\          
        $^2$ University of Lorraine, CNRS, Inria, Loria, 5400, Nancy, France,\\ \{romain.serizel\}@loria.fr\\ 
 }

\begin{document}

\ninept
\maketitle

\begin{sloppy}

\begin{abstract}
Deep learning has been widely used recently for sound event detection and classification. Its success is linked to the availability of sufficiently large datasets, possibly with corresponding annotations when supervised learning is considered. In bioacoustic applications, most tasks come with few labelled training data, because annotating long recordings is time consuming and costly. Therefore supervised learning is not the best suited approach to solve bioacoustic tasks. The bioacoustic community recasted the problem of sound event detection within the framework of few-shot learning, i.e. training a system with only few labeled examples. The few-shot bioacoustic sound event detection task in the DCASE challenge focuses on detecting events in long audio recordings given only five annotated examples for each class of interest. In this paper, we show that learning a rich feature extractor from scratch can be achieved by leveraging data augmentation using a supervised contrastive learning framework. We highlight the ability of this framework to transfer well for five-shot event detection on previously unseen classes in the training data. We obtain an F-score of 63.46\% on the validation set and 42.7\% on the test set, ranking second in the DCASE challenge. We provide an ablation study for the critical choices of data augmentation techniques as well as for the learning strategy applied on the training set. Our code is available on Github.\footnote{ : \url{https://github.com/ilyassmoummad/dcase23_task5_scl}}
\end{abstract}

\begin{keywords}
Contrastive learning, representation learning, transfer learning, few-shot learning, bioacoustic sound event detection.
\end{keywords}

\section{Introduction}
\label{sec:intro}

\begin{figure*}[]
\begin{minipage}[b]{1.0\linewidth}
  \centering
  \centerline{\includegraphics[width=1.\textwidth]{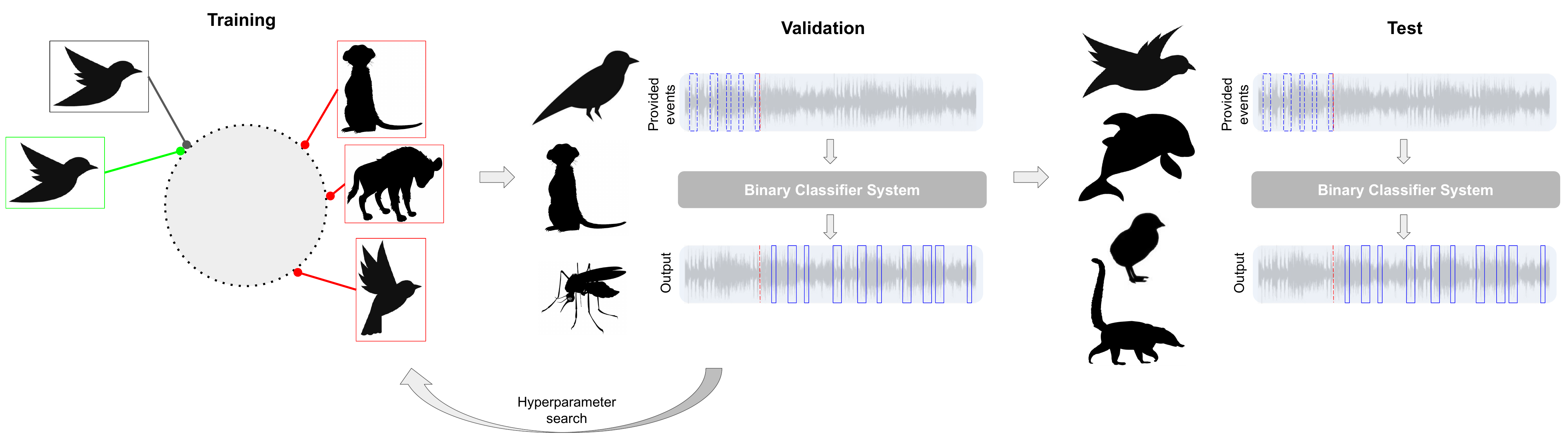}}
\end{minipage}
\caption{Overview of the proposed framework}
\label{fig:pip}
\end{figure*}

Sound Event Detection (SED) is the task of recognizing sound events, including determining their onsets and offsets, as well as recognizing them. SED has many applications in bioacoustics such as monitoring of biodiversity, studying animal behavior and identifying species. Automatic bioacoustic SED provides significant value in our understanding of animal populations and their interactions, as well as individuals and their behaviors. 
Standard SED systems leverage supervised learning as well as semi-supervised learning (DCASE Challenge Task 4) and have shown strong results in the recent years~\cite{survey1, survey2, serizelsed}. Numerous works focused on bird vocalization due to availability of large bird sound datasets~\cite{stowellbird, birddcase}. BirdNet~\cite{BirdNet} is a notable work for bird monitoring, able to identify nearly one thousand bird species. The approach involves training a model in a supervised fashion using a vast dataset comprising over one million labeled bird recordings, using extensive data-preprocessing and data augmentation techniques. 

However, such a large scale data collection for training systems is not always feasible in bioacoustics. The challenge lies not only in obtaining annotations but also in acquiring the audio samples themselves (e.g. for rare species or fields that are difficult to reach).
As a consequence, bioacoustics SED is considered as a collection of numerous small-data problems, each requiring specialized systems for their individual solutions. Thus, the community of bioacoustics recasted bioacoustic SED as a few-shot learning (FSL) problem~\cite{fssed, dcase22t5}.

FSL is a machine learning problem where a model has to learn to adapt to new classes of data unseen during training with only few labeled samples. FSL is adapted for many applications where acquisition or annotation is expensive or time consuming. The annual challenge on detection and classification of acoustic scenes and events (DCASE) organized a third edition for the task of few-shot bioacoustic sound event detection. This task focuses on SED in a FSL setting for mammal and bird vocalizations. The goal is to create a system that learns from five exemplar vocalizations (shots) to detect instances of these vocalizations in test audio recording.

Prototypical networks (ProtoNets)~\cite{protonet} were proposed as a baseline to solve the FSL problem of detecting animal sound events in the DCASE challenge~\cite{dcase22t5}. ProtoNets, a meta-learning framework, have been state-of-the-art FSL audio systems in the recent years~\cite{metaaudio, fsd-fs}. The goal of meta-learning training is to develop models that can quickly adapt to new tasks with minimal data by simulating the test scenario within the training process. In Computer Vision, simple transfer learning methods have been shown to outperform meta-learning methods in FSL~\cite{tianfs, easy} in several datasets such as MiniImageNet and TieredImagenet, in which case the domain shift between the training data and the few shot generalization is small enough. Here, we propose to test transfer learning to solve FSL problems for the bioacoustic SED~\cite{dcase22t5}.

As the generalization capability of the feature extractor is crucial for efficient transfer learning, we propose to train a model using the supervised contrastive learning framework (SCL)~\cite{scl}. Numerous contrastive learning methods have been proposed in the self-supervised learning (SSL) literature~\cite{simclr, swav, uclser}, but the fundamental concept of pulling together positive pairs and pushing apart negative pairs remains the same across these approaches. The positive pairs consist of similar samples, while negative pairs consist of dissimilar samples. The selection of these pairs can be achieved through various means, such as data augmentation techniques~\cite{simclr} and/or utilizing class labels as in done in SCL~\cite{scl}. The representations learned using this framework have shown competitive transfer learning performance with SSL and cross-entropy (CE) learning on a variety of downstream tasks in vision~\cite{scl}. In audio, the works of Moummad et al.~\cite{mscl} and Nasiri et al.~\cite{soundclr} have demonstrated strong generalization capabilities of SCL. 

Following the training of the feature extractor using SCL on the training set, the learned model is transferred to the validation set to conduct hyperparameter search. The optimal hyperparameter setting determined from this process is then employed on the test set for evaluation.
In summary, our contribution revolves around the proposition of employing supervised contrastive learning to train a feature extractor that can be transferred to new few-shot bioacoustic sound event detection tasks.



\section{METHOD}
\label{sec:method}

This section provides a comprehensive overview of the methodology employed in this study (Figure~\ref{fig:pip}). Firstly, we present the SCL framework utilized for pre-training a good feature extractor model. Secondly, we describe the data augmentation techniques employed to enhance the diversity and robustness of the learned features. Finally, we detail our transfer learning strategy for adapting the pre-trained model to effectively tackle novel tasks.

\subsection{Supervised Contrastive Learning}
\label{sec:scl}

SCL consists in learning an embedding space in which the samples with the same class labels are close to each other, and the samples with different class labels are far from each other. Formally, a composition of an encoder $f$ and a shallow neural network $h$ called a projector (usually a MLP with one hidden layer) are trained to minimize the distances between representations of samples of the same class while maximizing the distances between representations of samples belonging to different class.  After convergence, $h$ is discarded, and the encoder $f$ is used for transfer learning on downstream tasks. The supervised contrastive loss (SCL) is calculated as follows:

\begin{equation}
\label{scl}
\mathcal{L}^{SCL} = \sum_{i\in I}\frac{-1}{|P(i)|}\sum_{p\in P(i)}\log{\frac{\text{exp}\left(\boldsymbol{z}_i\boldsymbol{\cdot}\boldsymbol{z}_p/\tau\right)}{\sum\limits_{s\in S(i)}\text{exp}\left(\boldsymbol{z}_i\boldsymbol{\cdot}\boldsymbol{z}_s/\tau\right)}}
\end{equation}
where $i\in I=\{1...2 N\}$ is the index of an augmented sample within a training batch, containing two views of each original sample. These views are constructed by applying a data augmentation function $A$ twice to the original samples. $\boldsymbol{z}_{i}=h(f(A(\boldsymbol{x}_{i})))\in\mathbb{R}^{D_P}$ where ${D_P}$ is the projector's dimension. ${P(i)={\{p\in I:{{y}}_p={{y}}_i}\}}$ is the set of indices of all positives in the two-views batch distinct from $i$ sharing similar label with $i$. $|P(i)|$ is its cardinality, ${S(i)={\{s\in I:{s}\neq{i}}\}}$, the $\boldsymbol{\cdot}$ symbol denotes the dot product, and $\tau\in\mathbb{R}^{+*}$ is a scalar temperature parameter that controls the penalty strength on hard negative samples.

\subsection{Data Augmentation}
\label{sec:da}

Data augmentation is crucial for learning a good feature extractor as advocated by the SSL literature~\cite{simclr, scl, swav, nnclr}. To this end, we adopt several augmentation modules derived from the audio representation learning domain~\cite{uclser,luyuwang,byola}. The following augmentations are sequentially applied in the prescribed order and are iteratively employed twice on the same data, with the exception of spectrogram mixing, which is exclusively applied to a single view (based on our experimental findings, this configuration demonstrated superior performance). To demonstrate the significance of each augmentation technique, an ablation study is conducted in the subsequent section.

\begin{itemize}
    \item[--] Spectrogram mixing: we add background sounds using random samples from the same batch. The mixing follows:
    $\hat{x_1} = \alpha x_1 + (1-\alpha) x_2$.
    where $\hat{x_1}$ is considered as a view of $x_1 $ and $x_2$ is a random sample from the batch.
    \item[--] Frequency shift: we approximate frequency shift by shifting the spectrogram upwards by few bands.
    \item[--] Random crop: we crop a patch from the spectrogram along the time axis.
    \item[--] Spectrogram resize: this augmentation is applied after the crop to restore the spectrogram to its original size.
    \item[--] Power gain: we attenuate the power of the spectrogram by multiplying it with a coefficient sampled uniformly between 0.75 and 1. 
    \item[--] Additive white Gaussian noise: we add a small additive white Gaussian noise to the view
\end{itemize}

\subsection{Transfer Learning}
\label{sec:tl}

After training the feature extractor, we transfer the model to the validation and test tasks. Each audio file is treated independently as a separate SED problem (as the challenge rules specify). We extract the features of the five positive annotated prototypes (shots) indicating the occurrence of the event of interest. We select intervals preceding the positive events as for the negative prototypes indicating the absence of the event. We train a binary classifier on these two prototypes using cross-entropy loss. The encoder layers can be either frozen or fine-tuned. We use a sliding window along the audio file (starting from the end of the fifth positive shot) to select queries for making predictions. The class activity is determined independently in each query window using the classifier. The onsets and offsets decision is made based on the precise moment when the label for the window transitions from a negative class to a positive class and from a positive class to a negative class, respectively.

\section{Experiments}
\label{sec:exp}

\subsection{Data}
\label{sec:data}

The bioacoustic few-shot sound event detection DCASE task development set consists of a training set and a validation set, for more details we refer the reader to the description of the task in 2022~\cite{dcase22t5} as these sets did not change from the previous edition.

\subsubsection{Training}
\label{sec:train}

We train our system on the official training set. We select all the positively annotated segments within each audio file. We compute Mel spectrogram features with a FFT of size 512, a hop length of 128, a number of mels of 128 and a sampling rate of 22.05~kHz. Each positive annotated segment from the training set is chunked into patches of length 200~ms with 100~ms overlap. We apply min-max normalization on each patch.

\subsubsection{Validation and test}
\label{sec:val}

For each audio file, we extract the first five positively annotated segments. The duration of these segments varies due to the wide range of animals and classes covered by the dataset. Following the approach proposed by Tang et al.~\cite{tang}, we determine the window length based on the mean duration of the events in the file. To compute Mel spectrogram features, we employ identical parameters and normalization technique as those used during the training phase. The shift size equals to half of the window length to predict the class for each query window along the remaining duration of the audio. 

\subsection{Model}
\label{sec:model}

We use a ResNet~\cite{resnet} consisting of three blocks, each comprising three convolutional layers. The feature maps of these convolutions have sizes of 64, 128, and 256, respectively. Following each convolutional layer, we apply batch normalization and a leaky rectified linear unit (ReLU) activation function. Max pooling operations are performed after each block. Specifically, we employ a 2x2 kernel for the first and second blocks, while for the third block, we use a 1x2 kernel. This choice is made to preserve frequency information by avoiding excessive pooling of the frequency bands, as suggested by Hertkorn~\cite{hertkorn}.

To ensure consistent output dimensions despite varying input lengths, we incorporate adaptive max pooling at the end of the network. This pooling operation is configured to yield a desired output size of (8, 1), resulting in a latent vector of size 8 x 256 = 2048. A MLP projector is added, consisted of a hidden layer with a dimension of 2048 and an output layer with a dimension of 512.

\subsection{Training details}
\label{sec:td}

\subsubsection{Data augmentation}
\label{sec:dad}
The spectrogram mixing coefficient $\alpha$ is sampled from a $\beta(5,2)$ distribution. The frequency shift size is uniformly sampled between 0 and 10. The crop size (i.e. how much total duration is kept from the original audio) in the Random crop augmentation is uniformly sampled between 60\% and 100\%. Power gain augmentation is achieved by multiplying the mel spectrogram with a coefficient uniformly sampled between 0.75 and 1. The additive white Gaussian noise is incorporated by adding noise with a mean of zero and a variable standard deviation, which is uniformly chosen between 0 and 0.1.

\subsubsection{Training and evaluation}
\label{sec:te}

\begin{table*}[t]
\centering
\caption{Performance of different systems on the validation set; freezing all layers, fine-tuning one, two or all three layers.}
\label{Tab:perfv}
\scalebox{1}{%
\begin{tabular}{l|c|c|c|ccc|ccc|ccc}
\hline
\multirow{2}{*}{System} & \multirow{2}{*}{Precision} & \multirow{2}{*}{Recall} & \multirow{2}{*}{F1-score} & \multicolumn{3}{c|}{HB}                                         & \multicolumn{3}{c|}{ME}                                         & \multicolumn{3}{c}{PB}                                          \\ \cline{5-13} 
                        &                            &                         &                           & \multicolumn{1}{c|}{Pr}    & \multicolumn{1}{c|}{Re}    & F1    & \multicolumn{1}{c|}{Pr}    & \multicolumn{1}{c|}{Re}    & F1    & \multicolumn{1}{c|}{Pr}    & \multicolumn{1}{c|}{Re}    & F1    \\ \hline
Frozen           & 71.41                      & 55.19                   & 62.26                     & \multicolumn{1}{c|}{77.14} & \multicolumn{1}{c|}{81.57} & 79.29 & \multicolumn{1}{c|}{65.45} & \multicolumn{1}{c|}{69.23} & 67.28 & \multicolumn{1}{c|}{72.64} & \multicolumn{1}{c|}{\textbf{36.17}} & \textbf{48.29} \\
FineTune-1           & \textbf{73.93}                      & \textbf{55.59}                   & \textbf{63.46}                     & \multicolumn{1}{c|}{\textbf{82.95}} & \multicolumn{1}{c|}{82.32} & 82.63 & \multicolumn{1}{c|}{67.69} & \multicolumn{1}{c|}{84.61} & 75.21 & \multicolumn{1}{c|}{\textbf{72.72}} & \multicolumn{1}{c|}{33.33} & 45.71 \\
FineTune-2           & 72.90                      & 55.14                   & 62.79                     & \multicolumn{1}{c|}{79.73} & \multicolumn{1}{c|}{89.72} & 84.43 & \multicolumn{1}{c|}{\textbf{74.60}} & \multicolumn{1}{c|}{\textbf{90.38}} & \textbf{81.73} & \multicolumn{1}{c|}{65.57} & \multicolumn{1}{c|}{31.06} & 42.19 \\
FineTune-all           & 67.08                      & 51.58                   & 58.32                     & \multicolumn{1}{c|}{81.20} & \multicolumn{1}{c|}{\textbf{91.38}} & \textbf{85.99} & \multicolumn{1}{c|}{58.75} & \multicolumn{1}{c|}{\textbf{90.38}} & 71.21 & \multicolumn{1}{c|}{65.00} & \multicolumn{1}{c|}{27.65} & 38.80 \\ \hline
\multicolumn{13}{c}{\scriptsize *We highlight in bold the best scores for each metric}
\end{tabular}
}
\end{table*}

We train our model from scratch on the training set using SCL framework with a temperature $\tau$ = 0.06 using SGD optimizer with a batch size of 128, a learning rate of 0.01 with a cosine decay schedule, momentum of 0.9, and a weight decay of 0.0001 for 50 epochs. After training, we discard the MLP projector and transfer the encoder to the validation and test sets by training a linear binary classifier on the pretrained representations. In this phase we used random resized crop along the time axis with a crop size ranging from 90\% to 100\% of the original size. We submitted four distincts systems to the challenge : freezing all pretrained layers (Frozen), or finetuning the last, two last and all layers (FineTune-1, FineTune-2 and Finetune-3). We optimize our systems using Adam optimizer with a learning rate of 0.01 for 20 epochs for the first system, and 40 epochs with a learning rate of 0.001 for the others. The selection of these hyperparameters is based on evaluation conducted on the validation set.

\subsection{Results}
\label{sec:res}

The performance of our four systems on the validation set is presented in Table~\ref{Tab:perfv}. For PB dataset, where events are short (therefore only few patches are available, because we divide longer events into multiple chunks), the first system outperforms the others, indicating that fine-tuning degrades the performance when only few positive patches are present. Conversely, for the HB dataset, where events tend to be longer, the third and fourth systems outperform the others. This indicates that finetuning a greater number of layers is advantageous when more positive patches are present. The second system demonstrates satisfactory performance across all datasets, outperforming the other systems across all datasets with a max F1 score of 63.46\%.
It is important to note that our results on the validation set exhibit significant variability, primarily attributed to the instability of our proposed cross-entropy adaptation strategy. We acknowledge this limitation and plan to address it in future work.

\begin{table}[h]
\centering
\caption{F-score on the test sets of the different submissions}
\label{Tab:perft}
\begin{tabular}{l|c}
\hline
 & F-score \\ 
\hline
Frozen & 35.6\% (35.3 - 36.0)                                                            \\
FineTune-1 & \textbf{42.7\% (42.2 - 43.1)}                                                            \\
FineTune-2 & 38.3\% (37.9 - 38.7)                                                            \\
FineTune-all & 34.4\% (33.9 - 34.8)                                                            \\ \hline
\multicolumn{2}{c}{\scriptsize *with 95\% confidence interval}
\end{tabular}
\end{table}

Table~\ref{Tab:perft} displays the performance scores of our systems on the test sets. Remarkably, the ranking order of these systems on the test set aligns with that observed on the validation set. This consistency further validates the robustness and generalizability of our models across different datasets.

\subsection{Ablation study}
\label{sec:abla}

Table~\ref{Tab:daa} presents our ablation study on data augmentation. Additionally, Table~\ref{Tab:met} compares pre-training methods : SCL, cross-entropy training (CE), and the self-supervised training method SimCLR~\cite{simclr}, which has the same formula as SCL but without positive label pairs. We perform these studies on the validation set using the first system Frozen, where we freeze all layers, as it better captures the impact of the pre-training strategy. We use the same hyperparameter setting described in~\ref{sec:td} for all experiments except for CE training where we use a learning rate of 0.0001 after thorough exploration. Additionally, we modify the training duration for SimCLR, extending it to 100 epochs. This adjustment is made to account for the longer training requirements typically associated with self-supervised approaches. To ensure reliable results, we trained the model five times on the training set and conducted five evaluations for each run, resulting in a total of twenty-five runs per experiment.

\begin{table}[h]
\centering
\caption{Ablation of Data augmentation on the validation set}
\label{Tab:daa}
\begin{tabular}{l|c|c}
\hline
DA removed & Mean & [Min, Max] \\
\hline
Frozen (with all DAs) & 56.47 & [49.37, 62.39] \\
\hline
- Spectrogram mixing & 56.59 & [47.99, 64.65] \\
- Frequency shift & 58.60 & [49.73, 66.18]  \\
- Time stretch & 55.68 & [49.20, 62.83]  \\ 
- Power gain & 56.02 & [47.04, 63.01]  \\ 
- Additive noise & \textbf{59.04} & \textbf{[52.68, 67.47]}  \\ 
\hline
\multicolumn{3}{c}{\scriptsize *Best score is highlighted in bold}
\end{tabular}
\end{table}

The analysis presented in Table~\ref{Tab:daa} indicates that certain data augmentation techniques have a negative impact on the model's performance. Surprisingly, these effects were not evident during the challenge submission due to the limited number of experiments conducted at that time. Notably, the data augmentation setting that yielded the highest score was the setting without the additive white Gaussian noise to the spectrogram. This finding suggests that this particular augmentation strategy was either enforcing an invariance that is not beneficial for the downstream task at hand, or that the task becomes hard given the small size of the training dataset.

We can observe from the results of Table~\ref{Tab:met} that SCL consistently outperforms both SimCLR and CE frameworks for transfer learning. The superior performance of SCL highlights its efficacy in capturing discriminative features. These findings emphasize the importance of incorporating SCL as a powerful framework for advancing feature representation learning, particularly for enhancing transferability in downstream tasks.

\begin{table}[h]
\centering
\caption{Ablation of the pretraining methods on the validation set}
\label{Tab:met}
\begin{tabular}{c|c|c}
\hline
Method & Mean & [Min, Max] \\
\hline
CE & 51.96 & [43.013-57.42]  \\ 
SimCLR & 50.89 & [39.28-57.41]  \\ 
SCL & \textbf{56.27} & \textbf{[49.37, 62.39]}  \\ 
\hline
\multicolumn{3}{c}{\scriptsize *Highest F-score is highlighted in bold}
\end{tabular}
\end{table}

\section{Discussion and Perspectives}
\label{sec:disc}

In this study, we have provided a comprehensive description of a simple approach for bioacoustic few-shot sound event detection. We have detailed the methodology behind the systems we developed and submitted for the DCASE 2023 challenge task five. Our approach involves pretraining a feature extractor using supervised contrastive learning and data augmentation on the training set, followed by training binary classifiers on positive and negative prototypes for each audio file in the validation/evaluation sets. We proposed four systems. The first system, which utilized a linear classifier on frozen representations, demonstrated the robustness and transferability of the learned features. When fine-tuning the last layer (the second system) or the last two layers (the third system), the performance is increased. However, our current adaptation strategy, involving training classifiers on available shots, showed performance instability. We also note the gap in performance between the validation and the test sets. HB validation dataset is made of controlled lab recordings, which may make the detection easier, while PB recordings are in the wild with noisy background. Settings of the test set are more close to PB than HB~\cite{nolasco2023}.
To address the limitation and instability of our approach, future work will explore more effective adaptation techniques such as meta-learning. Notably, the winning systems in the 2022 and 2023 editions of the DCASE bioacoustic few-shot sound event detection challenge (Tang et al.~\cite{tang}; Du et al.~\cite{Du}) employed a frame-level approach, offering a higher time resolution capability compared to our window-level approach. Exploring the frame-level approach, as well as a proposal-based approach~\cite{proposal} for detecting variable length temporal regions of interest, which has not been previously investigated in this task, will be considered for future research. Combining representation learning (meta-learning, self-supervised learning, or supervised learning) is a promising direction for learning useful representation leveraging knowledge from large data, that can transfer well to new tasks.

\bibliographystyle{IEEEtran}
\bibliography{refs}

%
%
%
%
%
%
%
%
%

\end{sloppy}
\end{document}